\documentstyle[emulateapj]{article}

\submitted{to appear in the Astrophysical Journal, Letters}

\lefthead{Hachisu et al.}
\righthead{Light Curve of U Scorpii Quiescence}

\begin{document}

\title{A MODEL FOR THE QUIESCENT PHASE OF THE RECURRENT NOVA
U SCORPII}

\author{Izumi Hachisu}
\affil{Department of Earth Science and Astronomy, 
College of Arts and Sciences, University of Tokyo,
Komaba, Meguro-ku, Tokyo 153-8902, Japan; hachisu@chianti.c.u-tokyo.ac.jp}

\author{Mariko Kato}
\affil{Department of Astronomy, Keio University, 
Hiyoshi, Kouhoku-ku, Yokohama 223-8521, Japan; mariko@educ.cc.keio.ac.jp}

\author{Taichi Kato and Katsura Matsumoto}
\affil{Department of Astronomy, Kyoto University, 
Kitashirakawa, Sakyo-ku, Kyoto 606-8502, Japan; 
tkato@kusastro.kyoto-u.ac.jp, katsura@kusastro.kyoto-u.ac.jp}

\and 

\author{Ken'ichi Nomoto}
\affil{Department of Astronomy and Research Center for the Early 
Universe, University of Tokyo, 
Bunkyo-ku, Tokyo 113-0033, Japan \\ e-mail: 
nomoto@astron.s.u-tokyo.ac.jp}




\begin{abstract}
A theoretical light curve is constructed for the quiescent phase of
the recurrent nova U Scorpii in order to resolve the existing distance 
discrepancy between the outbursts ($d \sim 6$ kpc) 
and the quiescences ($d \sim 14$ kpc).
Our U Sco model consists of a very massive white dwarf (WD),
an accretion disk (ACDK) with a flaring-up rim, and  
a lobe-filling, slightly evolved, main-sequence star (MS).
The model properly includes an accretion luminosity of the WD, 
a viscous luminosity of the ACDK, a reflection effect 
of the MS and the ACDK irradiated by the WD photosphere.
The $B$ light curve is well reproduced by a model
of 1.37 $M_\odot$ WD $+$ 1.5 $M_\odot$ MS (0.8---2.0 $M_\odot$
MS is acceptable) with an ACDK having
a flaring-up rim, and the inclination angle 
of the orbit $i \sim 80 \arcdeg$.  The calculated color is rather 
blue ($B-V \sim 0.0$) for a suggested mass accretion rate
of $2.5 \times 10^{-7} M_\odot$ yr$^{-1}$, thus 
indicating a large color excess of $E(B-V) \sim 0.56$ 
with the observational color of $B-V = 0.56$ in quiescence.  
Such a large color excess corresponds to an absorption 
of $A_V \sim 1.8$ and $A_B \sim 2.3$, which reduces
the distance to 6---8 kpc.  This is in good agreement 
with the distance estimation of 4---6 kpc for the latest outburst.  
Such a large intrinsic absorption  is very consistent with the recently
detected period change of U Sco, which is indicating a mass 
outflow of $\sim 3 \times 10^{-7} M_\odot$ yr$^{-1}$
through the outer Lagrangian points in quiescence.
\end{abstract}




%

\section{INTRODUCTION}
      U Scorpii is one of the best observed recurrent 
novae, the outbursts of which were recorded
in 1863, 1906, 1936, 1979, 1987, and the latest in 1999. 
Especially, the 1999 outburst was well observed 
from the rising phase to the cooling phase 
by many observers (e.g., \cite{mun99}; \cite{kah99}; \cite{lep99}) 
including eclipses (Matsumoto, Kato, \& Hachisu 2000).  
Based on Matsumoto et al.'s (2000) observation,
Hachisu et al. (2000) have constructed a theoretical light-curve
model for the 1999 outburst of U Sco and obtained various 
physical parameters of the recurrent nova.
Their main results are summarized as follows: 
(1) A direct light-curve fitting of the 1999 outburst 
indicates a very massive white dwarf (WD) of
$M_{\rm WD}= 1.37 \pm 0.01 M_\odot$.
(2) The envelope mass at the optical maximum is estimated 
to be $\Delta M \sim 3 \times 10^{-6} M_\odot$. 
(3) Therefore, the mass accretion rate of the WD is 
$\dot M_{\rm acc} \sim 2.5 \times 10^{-7} M_\odot$ yr$^{-1}$
during the quiescent phase between 1987 and 1999.
(4) An optically thick wind blows from
the WD and plays a key role in determining the nova
duration because it reduces the envelope mass (\cite{kat94}).
About 60\% of the envelope mass is carried away in the wind, which
forms an expanding shell as observed in T Pyx (e.g., \cite{shr89}).
The residual 40\% ($1.2 \times 10^{-6} M_\odot$)
is added to the helium layer of the WD.
(5) As a result, the WD can grow in mass at an average
rate of $\sim 1 \times 10^{-7} M_\odot$ yr$^{-1}$.
\par
     The above physical pictures are exactly the same as proposed by
Hachisu et al. (1999b) as a progenitor system
of Type Ia supernovae (SNe Ia).  
However, the distance to U Sco is still controversial because
the direct light-curve fitting results in a relatively
short distance of $\sim 6$ kpc (\cite{hac2000}), which 
is incompatible with the distance of $\sim 14$ kpc 
at the quiescent phase (e.g., \cite{web87}; \cite{war95}; 
\cite{kah99}, for a summary).  If the distance of $\sim 14$ kpc 
is the case, it could be hardly
consistent with the results (1) to (5) mentioned above.
\par
     Our purpose in this Letter is to construct 
a light-curve model for the quiescent phase 
and to rectify the distance to U Sco.
Our numerical method to obtain light curves has been 
described both in Hachisu \& Kato (1999) to explain the second peak 
of T CrB outbursts and in Hachisu et al. (2000) to reproduce
the light curve for the 1999 outburst of U Sco.  
Therefore, we mention only new parts of our numerical method in \S 2.  
In \S 3, by directly fitting our theoretical light curve 
to the observations, we derive the distance to U Sco.
Discussions follow in \S 4, especially in relation to 
the recently detected orbital-period change of U Sco
and a systemic mass loss through the outer 
Lagrangian points.  We also discuss the relation 
to a progenitor system of SNe Ia.

\section{THEORETICAL LIGHT CURVES}
\begin{figure}
\epsscale{.9}
\plotone{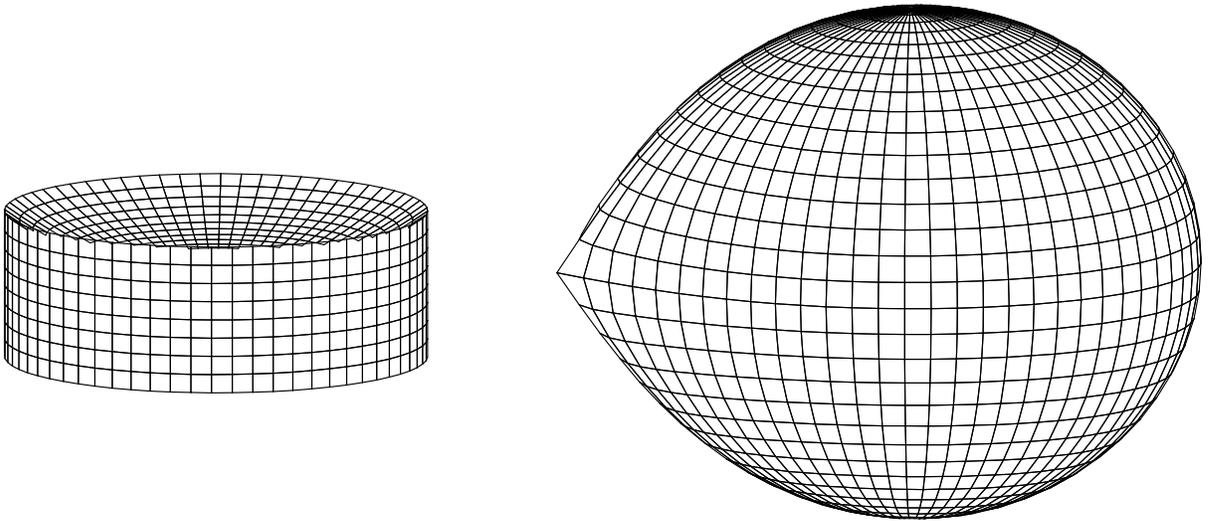}
\caption{
Configuration of our U Sco model at quiescent phase.
The cool component (right figure) is a slightly evolved 
MS ($1.5 M_\odot$) filling up its inner critical Roche lobe.  
Only the north and south polar areas of the secondary are 
slightly heated up by the hot component ($1.37 ~M_\odot$ WD, 
left figure) because a large part of the light from the hot 
component is blocked by the flaring-up edge of the accretion disk.
Both the hot component and the central part of the accretion disk
are not seen from the Earth because they are blocked 
by the flaring-up rim.  Here the separation is $a= 6.87 R_\odot$, 
the effective radii of the inner critical Roche lobes are
$R_1^*= 2.55 R_\odot$, and $R_2^*= R_2= 2.66 R_\odot$, 
for the primary WD and the secondary MS, respectively.
\label{uscofig_q35}}
\end{figure}

     Our U Sco model is graphically shown in Figure \ref{uscofig_q35}.  
Schaefer (1990) and Schaefer \& Ringwald (1995) observed 
eclipses of U Sco in the quiescent phase and determined 
the orbital period ($P= 1.23056$ days) and 
the ephemeris (HJD 2,451,235.777 $+$ 1.23056$E$) at the
epoch of mid-eclipse.  Thus, the companion is a main-sequence star 
(MS) which expands to fill its Roche lobe after a most
part of the central hydrogen is consumed.  We call such a star
"a slightly evolved" MS.  The inclination angle 
of the orbit ($i \sim 80\arcdeg$) is a parameter for fitting.

\placefigure{uscofig_q35}

We have assumed that (1) $M_{\rm WD}= 1.37 M_\odot$, 
(2) the WD luminosity of
\begin{equation}
L_{\rm WD} = {1 \over 2} {{G M_{\rm WD} \dot M_{\rm acc}} 
\over {R_{\rm WD}}} + L_{\rm WD,0},
\label{accretion-luminosity}
\end{equation}
where the first term is the accretion luminosity 
(e.g., Starrfield, Sparks, \& Shaviv 1988) and
the second term $L_{\rm WD,0}$ is the intrinsic luminosity of the WD,
and $R_{\rm WD}= 0.0032 R_\odot$ the radius of 
the $1.37 M_\odot$ WD, and (3) a black-body photosphere of the WD.
The accretion luminosity is 
$\sim 1700 L_\odot$ for a suggested mass accretion rate of
$\dot M_{\rm acc} \sim 2.5 \times 10^{-7} M_\odot$ yr$^{-1}$.
Here, we assume $L_{\rm WD,0}=0$ because the nuclear luminosity
is smaller than the accretion luminosity for this accretion rate,
but we have examined other two cases of $L_{\rm WD,0}=2000$ and 
4000 $L_\odot$ and found no significant differences in the distance 
as shown below.   
We do not consider the limb-darkening effect for simplicity.
\par
     It is assumed that the companion star is synchronously rotating 
on a circular orbit and its surface fills the inner 
critical Roche lobe as shown in Figure \ref{uscofig_q35}.
We neglect both the limb-darkening effect and
the gravity-darkening effect of the companion star for simplicity.
Here, we assume a 50\% irradiation efficiency of the companion 
star ($\eta_{\rm ir,MS}=0.5$).
We have examined the dependence of the distance on 
the irradiation efficiency (i.e., $\eta_{\rm ir,MS}=0.25$ and 1.0) 
but found no significant differences in the distance as shown below. 
The non-irradiated photospheric temperature $T_{\rm ph, MS}$ 
of the companion star is a parameter for fitting.
The mass of the secondary is assumed to be
$M_{\rm MS}= 1.5 M_\odot$.
\par
     The size of the accretion disk is a parameter for fitting
and defined as
\begin{equation}
R_{\rm disk} = \alpha R_1^*,
\label{accretion-disk-size}
\end{equation}
where $\alpha$ is a numerical factor indicating the size of 
the accretion disk, and $R_1^*$ the effective radius of 
the inner critical Roche lobe for the WD component
(e.g., Eggleton 1983).
We also assume that the accretion disk is axisymmetric and has 
a thickness given by 
\begin{equation}
h = \beta R_{\rm disk} \left({{\varpi} 
\over {R_{\rm disk}}} \right)^{\nu},
\label{flaring-up-disk}
\end{equation}
where $h$ is the height of the surface from the equatorial plane,
$\varpi$ the distance on the equatorial plane from the center of the WD, 
$\nu$ the power of the surface shape, 
and $\beta$ a numerical factor showing the degree of thickness 
and also a parameter for fitting.
We adopt a $\varpi$-squared law ($\nu=2$) simply to mimic the 
flaring-up effect of the accretion disk rim 
(e.g., Schandl, Meyer-Hofmeister, \& Meyer 1997), and have examined
the dependence of the distance on the power ($\nu=1.25$ and 3.0) 
without finding any significant differences as shown below.
\par
     The surface of the accretion disk also 
absorbs photons from the WD photosphere 
and reemits with a black-body spectrum at a local temperature.
We assume a 50\% irradiation efficiency of the companion star, i.e.,
$\eta_{\rm ir,DK}=0.5$  (e.g., \cite{sch97}).  We have examined
other two cases of $\eta_{\rm ir,DK}=0.25$ and 1.0, and found 
no significant differences in the distance as shown below.
The non-irradiated temperature of the disk surface is assumed 
to be determined by the viscous heating of the standard accretion
disk model.  Then, the disk surface temperature is given by 
\begin{equation}
\sigma T_{\rm ph, disk}^4 = {{3 G M_{\rm WD} \dot M_{\rm acc}} 
\over {8 \pi \varpi^3}} + \eta_{\rm ir,DK} 
{{L_{\rm WD}} \over {4 \pi r^2}} \cos\theta,
\end{equation}
where 
$r$ the distance from the WD center, and $\cos\theta$ 
the incident angle of the surface (e.g., \cite{sch97}).  
The temperature of the disk rim is assumed to be 3000 K. 

\placefigure{bmag_bv_color_paper}
\placetable{tbl-1}

\section{RESULTS}
\begin{figure}
\epsscale{.9}
\plotone{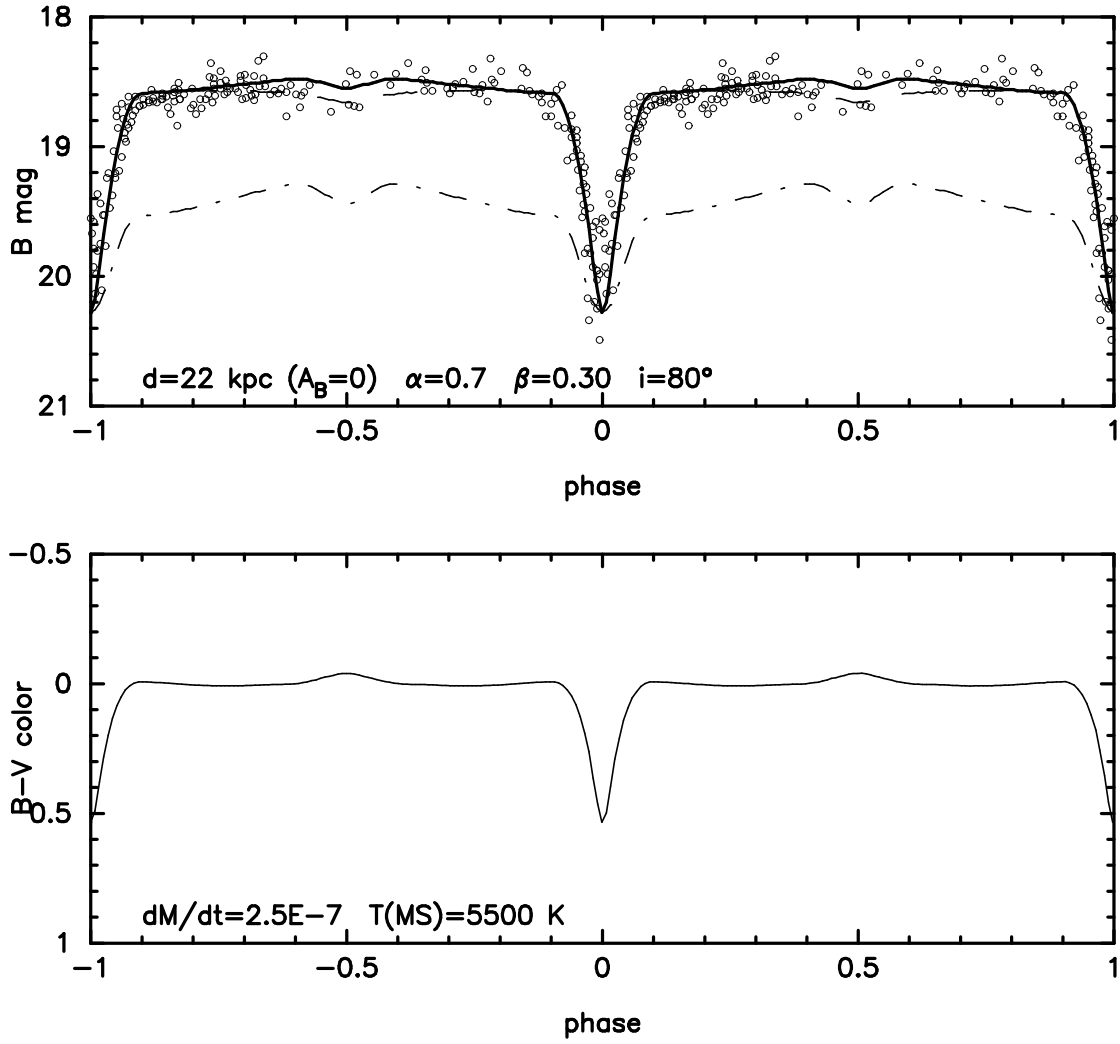}
\caption{
{\it Top}: theoretical $B$ light curves (thick solid line)
and {\it Bottom}: theoretical $B-V$ light curves (thin solid line)
plotted against the binary phase (two phases from $-1.0$ to $1.0$)
together with the observational points (open circles in the $B$ light
curve represent data taken from Schaefer 1990, 
but no data points in the $B-V$ light curve).
The model is a binary system of $1.37 M_\odot$ WD $+$ $1.5 M_\odot$ MS. 
The other parameters are printed in the figure. 
Two light curves are added in order to clarify each contribution 
of the light, for the case of no irradiation of the ACDK, 
$\eta_{\rm ir, DK}=0$ (dash-dotted), and for the case of 
no irradiation of the MS, $\eta_{\rm ir, MS}=0$ (dashed).
\label{bmag_bv_color_paper}}
\end{figure}

     Figure \ref{bmag_bv_color_paper} shows the observational 
points (open circles) by Schaefer (1990) together with 
our calculated $B$ light curve (thick solid line) for 
$\dot M_{\rm acc}= 2.5 \times 10^{-7} M_\odot$ yr$^{-1}$.
To fit our theoretical light curves with the 
observational points, we calculate $B$ light 
curves by changing the parameters of
$\alpha=0.5$---1.0 by 0.1 step, $\beta=0.05$---0.50 by 0.05 step,
$T_{\rm ph, MS}= 3500$---8000 K by 100 K step,
and $i=75$---$85\arcdeg$ by $1\arcdeg$ step and seek for
the best fit model.
The best fit parameters obtained are shown in Figure
\ref{bmag_bv_color_paper} (see also Table \ref{tbl-1}).  
\par
     There are five different contributions to the $B$-light 
($L_B$) in the system: the white dwarf ($L_{B1}$), 
the non-irradiated portions of 
the accretion disk ($L_{B2}$) and the donor star ($L_{B3}$), 
and the irradiated portions of the accretion disk ($L_{B4}$) 
and the donor star ($L_{B5}$).  In order to show each contribution,
we have added two light curves in Figure \ref{bmag_bv_color_paper}, 
that is, a non-irradiation case of the ACDK ($\eta_{\rm ir, DK}=0$, 
dash-dotted), and a non-irradiation case of the MS 
($\eta_{\rm ir, MS}=0$, dashed).  The light from the WD is 
completely blocked by the accretion disk rim, thus having no contribution,
$L_{B1}=0$.  The depth of the primary eclipse, 1.5 mag, means 
$L_{B3}= 0.25 L_{B}$ because the ACDK is completely occulted by the MS.  
The difference of 1 mag between the thick 
solid and dash-dotted lines indicates 
$L_{B4}=0.60 L_B$.  
The difference of 0.1 mag between the thick solid and dashed lines
indicates 
$L_{B5} = 0.10 L_B$.  Thus, we obtain each contribution: 
$L_{B1} = 0$, $L_{B2} = 0.05 L_B$, $L_{B3} = 0.25 L_B$,
$L_{B4} = 0.60 L_B$, and $L_{B5}= 0.10 L_B$.
\par
     Then we calculate the theoretical color index $(B-V)_c$ 
for these best fit models.  Here, we explain only the case of
$\dot M_{\rm acc}= 2.5 \times 10^{-7} M_\odot$ yr$^{-1}$.
By fitting, we obtain the apparent distance modulus 
of $m_{B, 0}= 16.71$, which corresponds to the distance 
of $d= 22$ kpc without absorption ($A_B=0$).  
On the other hand, we obtained a rather blue color index 
of $(B-V)_c= 0.0$ outside eclipses.  
Together with the observed color of $(B-V)_o= 0.56$ outside eclipses
(\cite{sch90}; \cite{sch95}), we derive a color excess
of $E(B-V)= (B-V)_o - (B-V)_c= 0.56$
Here, suffixes $c$ and $o$ represent the theoretically calculated 
and the observational values, respectively. 
Then, we expect an absorption of $A_V= 3.1 ~E(B-V)= 1.8$
and $A_B= A_V + E(B-V) = 2.3$.  
Thus, we are forced to have a rather short distance to U Sco of 7.5 kpc.
\par
     In our case of $\alpha=0.7$ and $\beta=0.30$, the accretion disk
is completely occulted at mid-eclipse.   The color index 
of $(B-V)_c= 0.53$ at mid-eclipse indicates
a spectral type of F8 for the cool component MS, 
which is in good agreement with the spectral type of F8$\pm$2 
suggested by Johnston \& Kulkarni (1992).
Hanes (1985) also suggested that a spectral type nearer F7 is preferred.
\par 
     For other mass accretion rates of $\dot M_{\rm acc}=$
(0.1---5.0)$\times 10^{-7} M_\odot$ yr$^{-1}$, we obtain
similar short distances to U Sco, as summarized in Table \ref{tbl-1}.  
It should be noted that, although the luminosity of the model
depends on our various assumptions of the irradiation efficiencies,
the $\varpi$-powered law of the disk, and the intrinsic luminosity
of the WD, the derived distance to U Sco itself is almost independent 
of these assumptions, as seen from Table \ref{tbl-2}.  
Therefore, the relatively short distance to U Sco ($\sim$ 6---8 
kpc) is a rather robust conclusion, at least, from the theoretical
point of view.

\placetable{tbl-2}

\section{DISCUSSION}
Matsumoto et al. (2000) observed a few eclipses during the 1999 
outburst and, for the first time, detected a significant period-change
of $\dot P / P = (-1.7 \pm 0.7) \times 10^{-6}$ yr$^{-1}$. 
If we assume the conservative mass transfer, 
this period change requires a mass transfer rate 
of $\gtrsim 10^{-6} M_\odot$ yr$^{-1}$ in quiescence.  
Such a mass transfer for 12 years is too high to be 
compatible with the envelope mass on the white dwarf, 
thus implying a non-conservative mass transfer in U Sco.
\par
     We have estimated the mass transfer rate for a non-conservative 
case by assuming that matter is escaping from the outer Lagrangian 
points and thus the specific angular momentum of the escaping matter 
is $1.7 a^2 \Omega_{\rm orb}$ (\cite{saw84}; \cite{hac99a}), 
where $a$ is the separation and $\Omega_{\rm orb} \equiv 2 \pi /P$.
Then the mass transfer rate from the companion is 
$\dot M_{\rm MS}= (-5.5 \pm 1.5) \times 10^{-7} M_\odot$ yr$^{-1}$ 
for $M_{\rm MS}= 0.8$---2.0 $M_\odot$ 
under the assumption that the WD receives matter at a rate of 
$\dot M_{\rm acc} = 2.5 \times 10^{-7} M_\odot$ yr$^{-1}$. 
The residual ($\sim 3 \times 10^{-7} M_\odot$ yr$^{-1}$),
which is escaping from the system, forms an excretion disk outside
the orbit of the binary.  Such an extended excretion disk/torus 
may cause a large color excess of $E(B-V)= 0.56$.  
\par
     Kahabka et al. (1999) reported the hydrogen column density of
(3.1---4.8)$\times 10^{21}$ cm$^{-2}$, which is much
larger than the Galactic absorption in the direction of U Sco
(1.4$\times 10^{21}$ cm$^{-2}$, \cite{dic90}), 
indicating a substantial intrinsic absorption.
It should also be noted here that Barlow et al. (1981) estimated the 
absorption toward U Sco by three ways: (1) the Galactic absorption
in the direction of U Sco, $E(B-V) \sim 0.24$ and $A_V \sim 0.7$,
(2) the line ratio of He~II during the 1979 outburst 
($t \sim$ 12 days after maximum), 
$E(B-V) \sim 0.2$ and $A_V \sim 0.6$, and (3) the Balmer line ratio
during the 1979 outburst ($t \sim$ 33---34 days after maximum), 
$E(B-V) \sim 0.35$ and $A_V \sim 1.1$.  The last one is 
significantly larger than the other two estimates.  
They suggested the breakdown of their case B
approximation in high density regions.  However, we may point out
another possibility that the systemic mass outflow from 
the binary system has already begun at $t \sim$ 33 days and, 
as a result, an intrinsic absorption is gradually increasing.
\par
     The mass of the companion star can be constrained from the mass
transfer rate.  Such a high transfer rate as $\dot M_{\rm MS} \sim 5.5
\times 10^{-7} M_\odot$ yr$^{-1}$ strongly indicates 
a thermally unstable mass transfer (e.g., \cite{heu92}), which  
is realized when the mass ratio is larger than 1.0---1.1, i.e., 
$q= M_{\rm MS}/ M_{\rm WD} >$ 1.0---1.1 for zero-age 
main-sequence stars (\cite{web85}).  This may pose a requirement 
$M_{\rm MS} \gtrsim 1.4 M_\odot$.  
We estimate the most likely companion mass of 1.4---1.6 $M_\odot$ 
from equation (11) in Hachisu et al. (1999b).
\par
     If the distance to U Sco is $\sim$ 6.0---8.0 kpc, 
it is located $\sim$ 2.3---3.0 kpc above the Galactic plane 
($b=22\arcdeg$).  The zero-age masses of the progenitor system to 
U Sco are rather massive (e.g., $8.0 ~M_\odot + 2.5 ~M_\odot$ 
from Hachisu et al. 1999b)
and it is unlikely that such massive stars were born 
in the halo.  Some normal B-type main-sequence stars have been found 
in the halo (e.g., PG0009+036 is located $\sim$ 5 kpc below 
the Galactic disk, \cite{smt96}), which were ejected from 
the Galactic disk because of their relatively high moving velocities
$\sim$100---200 km s$^{-1}$.  
The radial velocity of U Sco is not known but it is suggested that 
the $\gamma$-velocity is $\sim$50---100 km s$^{-1}$ 
from the absorption line velocities (\cite{joh92}; \cite{sch95}).  
If so, it seems likely that U Sco was ejected from the Galactic
disk with a vertical velocity faster than $\sim 20$ km s$^{-1}$ 
and has reached at the present place within 
the main-sequence lifetimes of a $\sim 3.0 M_\odot$ star 
($\sim 3.5\times 10^8$ yr).  
\par
     Now, we can understand the current evolutionary status 
and a further evolution of U Sco system.
The white dwarf has a mass $1.37 \pm 0.01 M_\odot$.  It is very
likely that the WD has reached such a large mass by mass accretion.
In fact the WD is currently increasing the mass of the helium layer at a
rate of $\dot M_{\rm He} \sim 1.0 \times 10^{-7} M_\odot$ yr$^{-1}$
(\cite{hac2000}).
We then predict that the WD will evolve as follows.  When the mass of
the helium layer reaches a critical mass after many cycles of 
recurrent nova outbursts, a helium shell flash will occur.  
Its strength is as weak as
those of AGB stars because of the high mass accretion rate
(\cite{nom82}).
A part of the helium layer will be blown off in
the wind, but virtually all of the helium layer will be burnt
into carbon-oxygen and accumulates in the white dwarf
(\cite{kat99h}).  Therefore, the WD mass can grow 
until an SN Ia explosion is triggered (\cite{nom84}).

\acknowledgments
We thank the anonymous referee for many critical comments 
to improve the manuscript.
This research has been supported in part by the Grant-in-Aid for
Scientific Research (07CE200, 08640321, 09640325, 11640226, 20283591) 
of the Japanese Ministry of Education, Science, Culture, and Sports.
KM has been financially supported as a Research Fellow 
for Young Scientists by the Japan Society for the Promotion of Science.

%
%


\clearpage
\begin{deluxetable}{ccccccccccc}
\footnotesize
\tablecaption{U Sco quiescent phase\tablenotemark{a}.
\label{tbl-1}}
\tablewidth{0pt}
\tablehead{
\colhead{$\dot M_{\rm acc}$} &
\colhead{$\alpha$} &
\colhead{$\beta$} &
\colhead{$T_{\rm ph, MS}$} &
\colhead{$m_{B,0}$} &
\colhead{outside eclipse}&
\colhead{mideclipse}&
\colhead{outside eclipse} &
\colhead{$A_V$} &
\colhead{$A_B$} &
\colhead{$d$} \nl
\colhead{($M_\odot$ yr$^{-1}$)} &
\colhead{} &
\colhead{} &
\colhead{(K)} &
\colhead{} &
\colhead{$(B-V)_c$} &
\colhead{$(B-V)_c$} &
\colhead{$E(B-V)$} &
\colhead{} &
\colhead{} &
\colhead{(kpc)}
} 
\startdata
5.0$\times 10^{-7}$ & 0.7 & 0.30 & 5900 & 17.18 & $-0.08$ & 0.45~(F5) 
& 0.64 & 2.01 & 2.65 & 8.0 \nl
2.5$\times 10^{-7}$ & 0.7 & 0.30 & 5500 & 16.71 & $+0.00$ & 0.53~(F8) 
& 0.56 & 1.76 & 2.32 & 7.5 \nl
1.0$\times 10^{-7}$ & 0.7 & 0.25 & 5000 & 16.02 & $+0.12$ & 0.66~(G4) 
& 0.44 & 1.38 & 1.82 & 6.9 \nl
5.0$\times 10^{-8}$ & 0.7 & 0.25 & 4600 & 15.45 & $+0.24$ & 0.78~(G9) 
& 0.32 & 1.01 & 1.33 & 6.7 \nl
2.5$\times 10^{-8}$ & 0.7 & 0.25 & 4200 & 14.72 & $+0.37$ & 0.91~(K2) 
& 0.19 & 0.60 & 0.79 & 6.1 \nl
1.0$\times 10^{-8}$ & 0.7 & 0.25 & 3700 & 13.58 & $+0.58$ & 1.13~(K5) 
& \nodata & \nodata & \nodata & 5.2 \nl
\enddata
\tablenotetext{a}{inclination angle $i=80\arcdeg$ for all cases
}
\end{deluxetable}

\begin{deluxetable}{lcccccccc}
\footnotesize
\tablecaption{Model dependence of the distance.
\label{tbl-2}}
\tablewidth{0pt}
\tablehead{
\colhead{model\tablenotemark{a}} &
\colhead{$T_{\rm ph, MS}$} &
\colhead{$m_{B,0}$} &
\colhead{outside eclipse}&
\colhead{mideclipse}&
\colhead{outside eclipse} &
\colhead{$A_V$} &
\colhead{$A_B$} &
\colhead{$d$} \nl
\colhead{} &
\colhead{(K)} &
\colhead{} &
\colhead{$(B-V)_c$} &
\colhead{$(B-V)_c$} &
\colhead{$E(B-V)$} &
\colhead{} &
\colhead{} &
\colhead{(kpc)}
} 
\startdata
$L_{\rm WD,0}=2000 L_\odot$ & 6100 & 17.31 & $-0.09$ & 0.41~(F3) 
& 0.65 & 2.04 & 2.60 & 8.4 \nl
$L_{\rm WD,0}=4000 L_\odot$ & 6400 & 17.59 & $-0.13$ & 0.36~(F1) 
& 0.69 & 2.17 & 2.87 & 8.8 \nl
$\eta_{\rm ir,MS}=1.0$ & 5500 & 16.76 & $+0.00$ & 0.53~(F8) 
& 0.56 & 1.76 & 2.32 & 7.7 \nl
$\eta_{\rm ir,MS}=0.25$ & 5500 & 16.66 & $+0.00$ & 0.53~(F8) 
& 0.56 & 1.76 & 2.32 & 7.4 \nl
$\eta_{\rm ir,DK}=1.0$ & 6000 & 17.16 & $-0.06$ & 0.43~(F4) 
& 0.62 & 1.95 & 2.57 & 8.3 \nl
$\eta_{\rm ir,DK}=0.25$ & 5300 & 16.39 & $+0.08$ & 0.58~(F9) 
& 0.48 & 1.51 & 1.99 & 7.6 \nl
$\nu=3.0$ & 5600 & 16.90 & $-0.04$ & 0.51~(F7) 
& 0.60 & 1.88 & 2.48 & 7.6 \nl
$\nu=1.25$ & 5300 & 16.39 & $+0.09$ & 0.58~(F9) 
& 0.47 & 1.47 & 1.95 & 7.7 \nl
\enddata
\tablenotetext{a}{$i=80\arcdeg$, 
$\dot M_{\rm acc}= 2.5 \times 10^{-7} M_\odot$ yr$^{-1}$, 
$\alpha=0.7$, $\beta=0.30$, $\nu=2$, $\eta_{\rm ir, MS}=0.5$,
$\eta_{\rm ir, DK}=0.5$, and $L_{\rm WD,0}=0$ are assumed, 
otherwise specified.
}
\end{deluxetable}

%
%
%
%

\clearpage


%

\end{document}